\documentclass[]{jetp}

\twocolumn

\usepackage[T1]{fontenc}
\usepackage[cp1251]{inputenc}  
\usepackage{babel}
\usepackage{graphicx}

\begin{document}

\English

\title{Hall Effect in Doped Mott -- Hubbard Insulator}

\rtitle{Hall Effect in Doped Mott -- Hubbard Insulator}

\affiliation{Institute for Electrophysics, RAS Ural Branch, Amundsen str. 106, 
Ekaterinburg 620016, Russia\\
$^*$II Physikalisches Institut, Universit\"at zu Koeln,
Zuelpicher Str. 77, 50937 Koeln, Germany}

\author {E.Z.} {Kuchinskii}
\author {N.A.} {Kuleeva}
\author {M.V.} {Sadovskii}
\author {D.I.} {Khomskii}


\abstract{We present theoretical analysis of Hall effect in doped Mott -- 
Hubbard insulator, considered as a prototype of cuprate superconductor.
We consider the standard Hubbard model within DMFT approximation.
As a typical case we consider the partially filled (hole doping) lower Hubbard
band. We calculate the doping dependence of both the Hall coefficient and Hall 
number and determine the value of carrier concentration, where Hall effect 
changes its sign. 
We obtain a significant dependence of Hall effect parameters on temperature. 
Disorder effects are taken into account in a qualitative way.
We also perform a comparison of our theoretical results with some known experiments
on doping dependence of Hall number in the normal state of YBCO and Nd-LSCO, 
demonstrating rather satisfactory agreement of theory and experiment.
Thus the doping dependence of Hall effect parameters obtained within Hubbard 
model can be considered as an alternative to a popular model of the quantum
critical point.}


\maketitle

\section{Introduction}

The studies of Hall effect in high -- temperature superconductors continues for
a long time. The early experiments demonstrated the significant dependence of
Hall effect parameters on temperature and doping, which are qualitatively
different from the case of ordinary metals \cite{Iye}. The complete understanding
of Hall effects in cuprates at present is absent.

In recent years much interest was attracted to experimental studies of Hall effect
at low temperatures in the normal state of high -- temperature superconductors
(cuprates), which is realized in very strong external magnetic fields
\cite{Boeb,Tal1,Tal2}. The observed anomalies of Hall effect in these 
experiments are usually attributed to reconstruction of Fermi surface due to
(antiferromagnetic) pseudogap formation and closeness to the corresponding
quantum critical point \cite{PrTal}.

Since the early days of theoretical studies of cuprates  the leading 
point of view is, that these systems are strongly correlated  and metallic 
(and superconducting) state is realized via doping of the initial Mott insulator
phase, which in a simplest case can be described within Hubbard model.
However, up to now there are only few papers where systematic studies of Hall 
effect dependence on doping and temperature were performed within this model
\cite{pruschke}.

Even the answer to a classical question on the doping level at which the Hall 
effect changes its sign is not perfectly clear . At small hole doping of an 
initial insulator, such as La$_2$CuO$_4$ or YBCO Hall effect is
obviously determined by hole concentration $\delta$. Then at what doping level
Hall effect changes its sign and when the transition from hole -- like Fermi
surface to electron -- like takes place? The answer to this question seems to
be important also for the general theory of transport phenomena in strongly
correlated systems. This paper is mainly devoted to the solution of this
problem.

\section{Hall conductivity and Hall coefficient}

One of the most general approaches to the studies of Hubbard model is the
dynamical mean field theory (DMFT) \cite{pruschke,georges96,Vollh10}, 
which gives an exact description of the system in the limit of infinite
spatial dimensions (lattice with infinite number of nearest neighbors).
General approaches allowing to overcome this rigid limitation are actively
developed \cite{GDMFT, RMP}, but as a rule these complicate the analysis very
much. In this paper we limit ourselves to the analysis of Hall effect within
the standard DMFT approximation.
The aim of this work is systematic study of concentration and temperature
dependence of Hall effect at different doping levels of the lower Hubbard band
and comparison of theoretical results with experiments on YBCO and
Nd-LSCO \cite{Tal1,Tal2}. Preliminary results were published in Ref. \cite{KKKS}.

In the standard DMFT \cite{pruschke,georges96,Vollh10} self -- energy of a
single -- electron Green's function  $G({\bf p}\varepsilon)$ is local,
i.e. independent of momentum. Due to this locality both the usual and Hall
conductivities are completely determined by the spectral density of this
Green's function
\begin{equation}
A({\bf p}\varepsilon)=-\frac{1}{\pi}ImG^R({\bf p}\varepsilon).
\label{SpDens}
\end{equation}
In particular, the usual (diagonal) static conductivity is given by 
\cite{pruschke}:
\begin{equation}
\sigma_{xx}=\frac{\pi e^2}{2\hbar a}\int_{-\infty}^{\infty}d\varepsilon
\left( -\frac{df(\varepsilon)}{d\varepsilon} \right)
\sum_{{\bf p}\sigma}\left( \frac{\partial \varepsilon ({\bf p})}{\partial p_x} \right) ^2
A^2({\bf p}\varepsilon), 
\label{Gxx}
\end{equation}
while Hall (non -- diagonal) conductivity is defined as:
\begin{eqnarray}
\sigma^H_{xy}=\frac{2\pi^2e^3aH}{3\hbar^2}\int_{-\infty}^{\infty}d\varepsilon
\left( \frac{df(\varepsilon)}{d\varepsilon} \right)\times\nonumber\\
\times\sum_{{\bf p}\sigma}\left( \frac{\partial \varepsilon ({\bf p})}{\partial p_x} \right) ^2
\frac{\partial^2 \varepsilon ({\bf p})}{\partial p_y^2}
A^3({\bf p}\varepsilon). 
\label{Gxy}
\end{eqnarray}
Here $a$ is the lattice parameter, $\varepsilon ({\bf p})$ is electron dispersion,
$f(\varepsilon)$ is Fermi distribution, and $H$ is magnetic field along $z$ axis. 
Then the Hall coefficient:
\begin{equation}
R_H=\frac{\sigma^H_{xy}}{H\sigma_{xx}^2}
\label{R_H}
\end{equation}
is also completely determined by the spectral density $A({\bf p}\varepsilon)$, 
which in the following will be calculated within the DMFT 
\cite{pruschke,georges96,Vollh10}.
Effective Anderson single -- impurity model of DMFT in this paper was solved with
numerical renormalization group (NRG) \cite{NRGrev}.

In the following we consider two basic models of the bare electron band.
The model with semi -- elliptic density of states (DOS) (per elementary cell
and single spin projection) is a reasonable 
approximation for three -- dimensional case:
\begin{equation}
N_0(\varepsilon)=\frac{2}{\pi D^2}\sqrt{D^2-\varepsilon^2},
\label{DOSd3}
\end{equation}
where $D$ is conduction band half -- width. We assume that the bare electronic
spectrum is isotropic. To find the momentum derivatives of the spectrum in this
model, entering Eqs. (\ref{Gxx}) and (\ref{Gxy}), we follow the procedure
proposed before in Ref. \cite{dis_hubb_2008}. Appropriate technical details are
presented in the Appendix.

For two -- dimensional systems, in anticipation of comparison with experimental
data for cuprates, we limit ourselves to the usual tight -- binding model
of electronic spectrum:
\begin{equation}
\varepsilon ({\bf p})=-2t(cos(p_xa)+cos(p_ya))-4t'cos(p_xa)cos(p_ya).
\label{SPtt'}
\end{equation}
Within this two -- dimensional model we consider a number of cases:\\ 
(1) spectrum with only nearest hoppings ($t'=0$) and full electron -- hole 
symmetry;\\
(2) spectrum with $t'/t=-0.25$, which qualitatively corresponds to electronic
dispersion in systems like LSCO;\\
(3) spectrum with $t'/t=-0.4$, which qualitatively corresponds to situation
observed in YBCO.

Below we present the results of detailed calculations of Hall coefficient for
all these models.

\section{Hall coefficient in two -- dimensional model of tight -- binding spectrum of electrons}

\begin{figure}
\includegraphics[clip=true,width=0.5\textwidth]{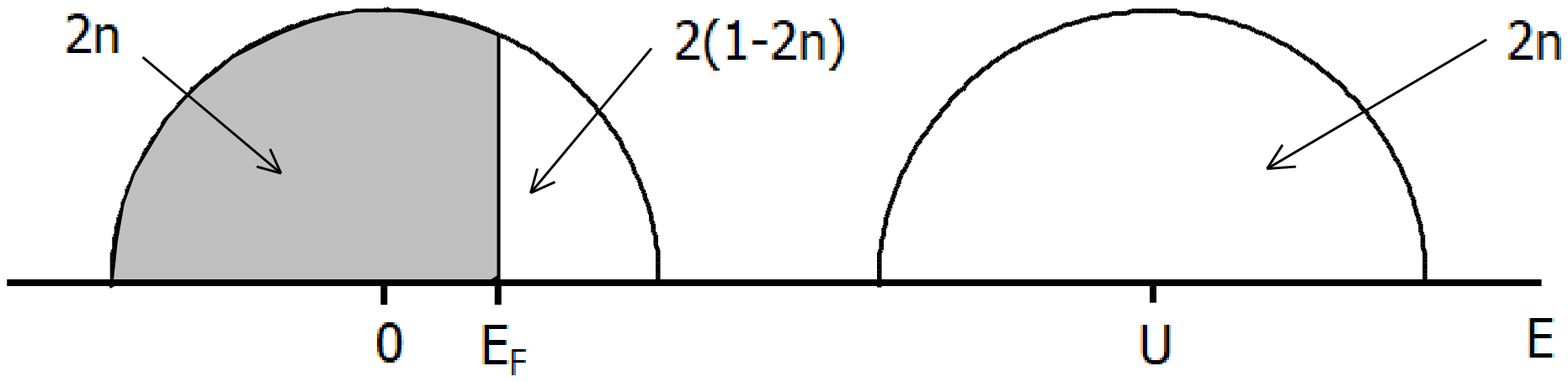}
\caption{Schematic picture of doping in Hubbard bands for the case of complete
electron -- hole symmetry.}
\label{fig_1a}
\end{figure}

Let us start with simplest qualitative analysis. It is easy to understand that
deep in Mott insulator state with well defined upper and lower Hubbard bands
the Hall coefficient under hole doping is in fact determined by filling the 
lower Hubbard band (the upper band is significantly higher in energy and is
practically unfilled). In this situation in the model with electron -- hole
symmetry (in two dimensions this corresponds to spectrum with $t'=0$), an 
estimate of band filling corresponding to the sign change of the Hall
coefficient can be obtained using very simple arguments.
Let us consider the paramagnetic phase with $n_{\uparrow}=n_{\downarrow}=n$, 
so that in the following $n$ denotes electron density per one spin projection,
so the the complete density of electrons is given by $2n$.  
Qualitatively the situation is illustrated in Fig. \ref{fig_1a}.
In lower Hubbard band (in the vicinity of the Fermi level $E=0$)
$2n$ electrons occupy states below the Fermi energy $E_F$.
An additional electron can go to the upper Hubbard band in the vicinity of
$E\sim U$, where we also have $2n$ states. It also can go to the lower Hubbard
band, where there still remain $2(1-2n)$ empty states in the region of $E>E_F$. 
Summing we obtain $2n+2(1-2n)+2n=2$ as it should be. The sign of the Hall
coefficient will change at the half filling of the lower band, when
$2n=2(1-2n)$. Now it is clear that the value of the critical concentration
is $n_c=1/3$.

The same result is easily obtained in Hubbard I approximation, where the
Green's function spin -- up electron takes the form \cite{Khom}:
\begin{equation}
G^R_{\uparrow}(\varepsilon {\bf p})=\frac{1-n_{\downarrow}}{\varepsilon-\varepsilon_-({\bf p})+i\delta}+
\frac{n_{\downarrow}}{\varepsilon-\varepsilon_+({\bf p})+i\delta}.
\label{HubbardI}
\end{equation}
where $\varepsilon_{\pm}(\bf p)$ is quasiparticle spectrum in upper and lower
Hubbard bands.
We see that in this approximation the number of states with upper spin
projection in the lower Hubbard band (first term in Eq. (\ref{HubbardI})) 
is in fact $1-n_{\downarrow}$. During hole doping of Mott insulator practically
all filling takes place in the lower Hubbard band, so that:
\begin{eqnarray}
&& n=n_{\uparrow}\approx\nonumber\\
&& \approx (1-n_{\downarrow})\int_{-\infty}^{\infty}d\varepsilon f(\varepsilon)
\left( -\frac{1}{\pi}Im\sum_{\bf p} \frac{1}{\varepsilon-\varepsilon_-({\bf p})+
i\delta}\right)=\nonumber\\
&& \equiv (1-n)n_0.
\label{H1}
\end{eqnarray}
Then at half -- filling of the lower Hubbard band we have $n_0=1/2$ and the
Hall coefficient (effective mass of the 
quasiparticles) changes its sign at $n=n_c=1/3$, corresponding to our
previous estimate.

\begin{figure}
\includegraphics[clip=true,width=0.6\textwidth]{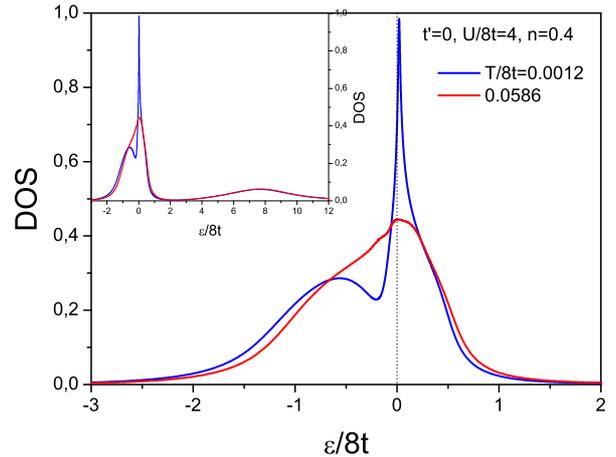}
\caption{Density of states (DOS) in doped Mott insulator at different
temperatures. Parameters of the Hubbard model are shown in the Figure,
$8t$ is the initial band -- width. At the insert --- 
the density of states in a wide energy interval including the upper Hubbard
band.}
\label{fig1}
\end{figure}

In general case situation is obviously more complicated. In strongly correlated
systems Hall coefficient (and other electronic properties) become significantly
dependent on temperature. At low temperature in these systems DMFT approximation
leads, besides the formation of lower and upper Hubbard bands, to the 
appearance of a narrow quasiparticle band, or quasiparticle peak in the density
of states \cite{pruschke,georges96,Vollh10}.
In hole doped Mott insulator (in the following we consider only hole doping)
such a peak appears close to the upper edge of the lower Hubbard band
(cf. Fig. \ref{fig1}). Thus at low temperatures the Hall coefficient is mainly
determined by filling of this quasiparticle band.
At high enough temperature (of the order or higher than quasiparticle peak
width) quasiparticle peak is damped and the Hall coefficient is mainly
determined by filling of the lower Hubbard band.
Thus, in general case it is necessary to consider two different temperature 
regimes for Hall coefficient.

\begin{figure}
\includegraphics[clip=true,width=0.5\textwidth]{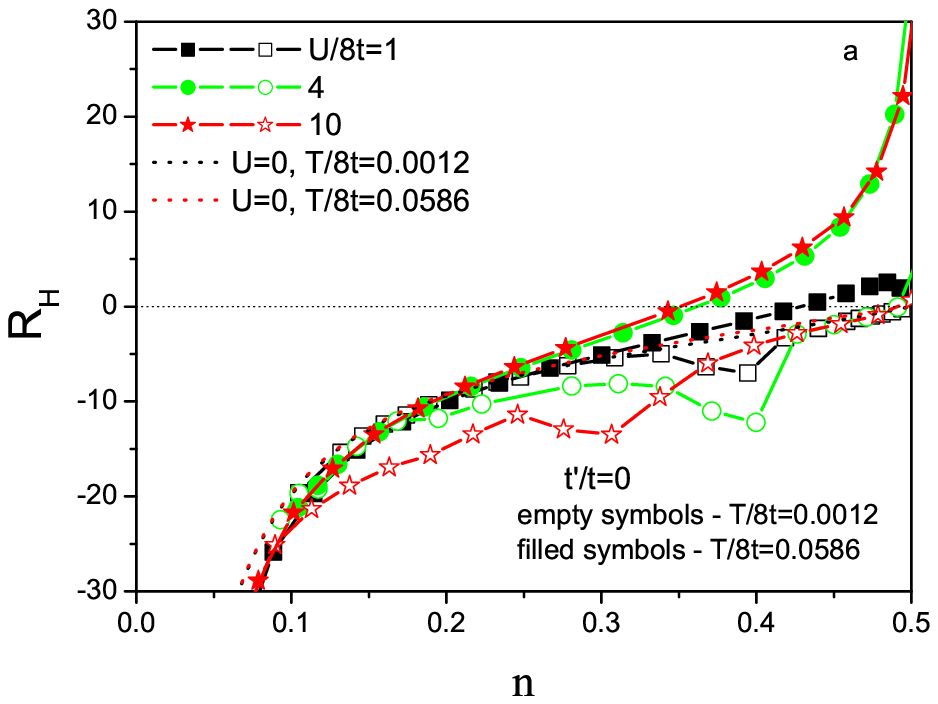}
\includegraphics[clip=true,width=0.5\textwidth]{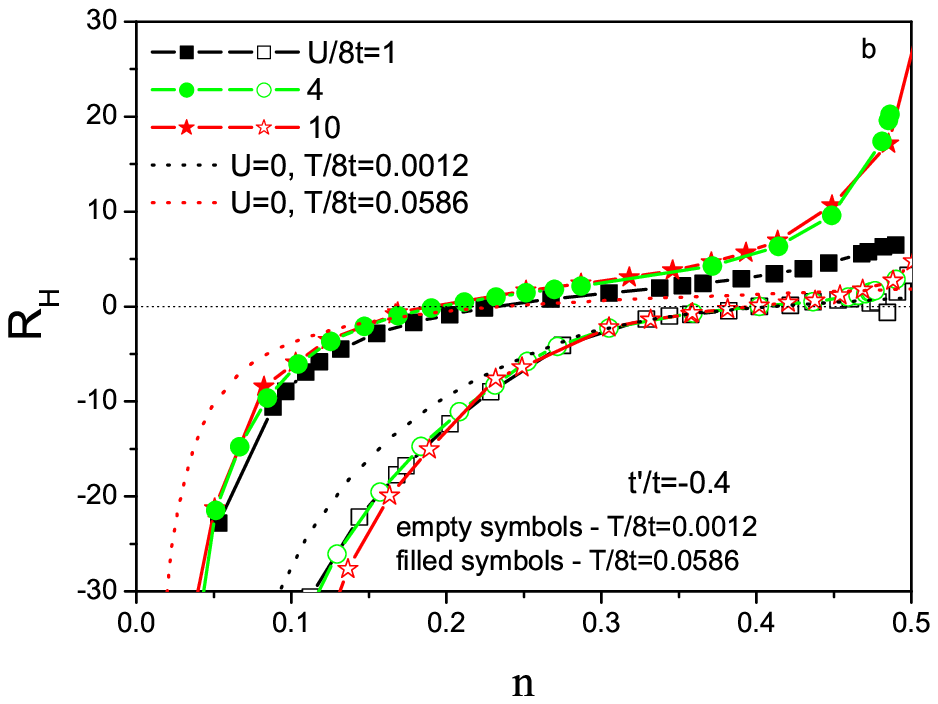}
\caption{Dependence of Hall coefficient on correlation strength $U$
 on band filling for $t'=0$ (a) and $t'/t=-0.4$ (b) in low temperature regime
(empty symbols) and in high temperature regime (filled symbols).}
\label{fig2}
\end{figure}

In low temperature regime both the width and the amplitude of quasiparticle 
peak depend on filling and temperature. Increasing temperature leads to
widening of quasiparticle peak and some shift of the Fermi level below the
maximum of this peak (cf. Fig. \ref{fig1}). This may lead to a significant
drop of the Hall coefficient, though further increase of the temperature
leading to the damping of the quasiparticle peak leads to the growth of this
coefficient. Thus the relevant dependence of the quasiparticle peak on 
band filling in the low temperature regime leads to the regions of non
monotonous filling dependence of the Hall coefficient (cf. Fig. \ref{fig2}(a)).

From Fig. \ref{fig2}(a) it is easy to see that high -- temperature behavior of
the Hall coefficient in doped Mott insulator ($U/2D=4;10$)  in a model with
full electron -- hole symmetry ($t'=0$), completely confirms the qualitative
estimate given above. However, this estimate becomes invalid in the case of
noticeable breaking of electron -- hole symmetry (cf. Fig. \ref{fig2}(b)).

\begin{figure}
\includegraphics[clip=true,width=0.5\textwidth]{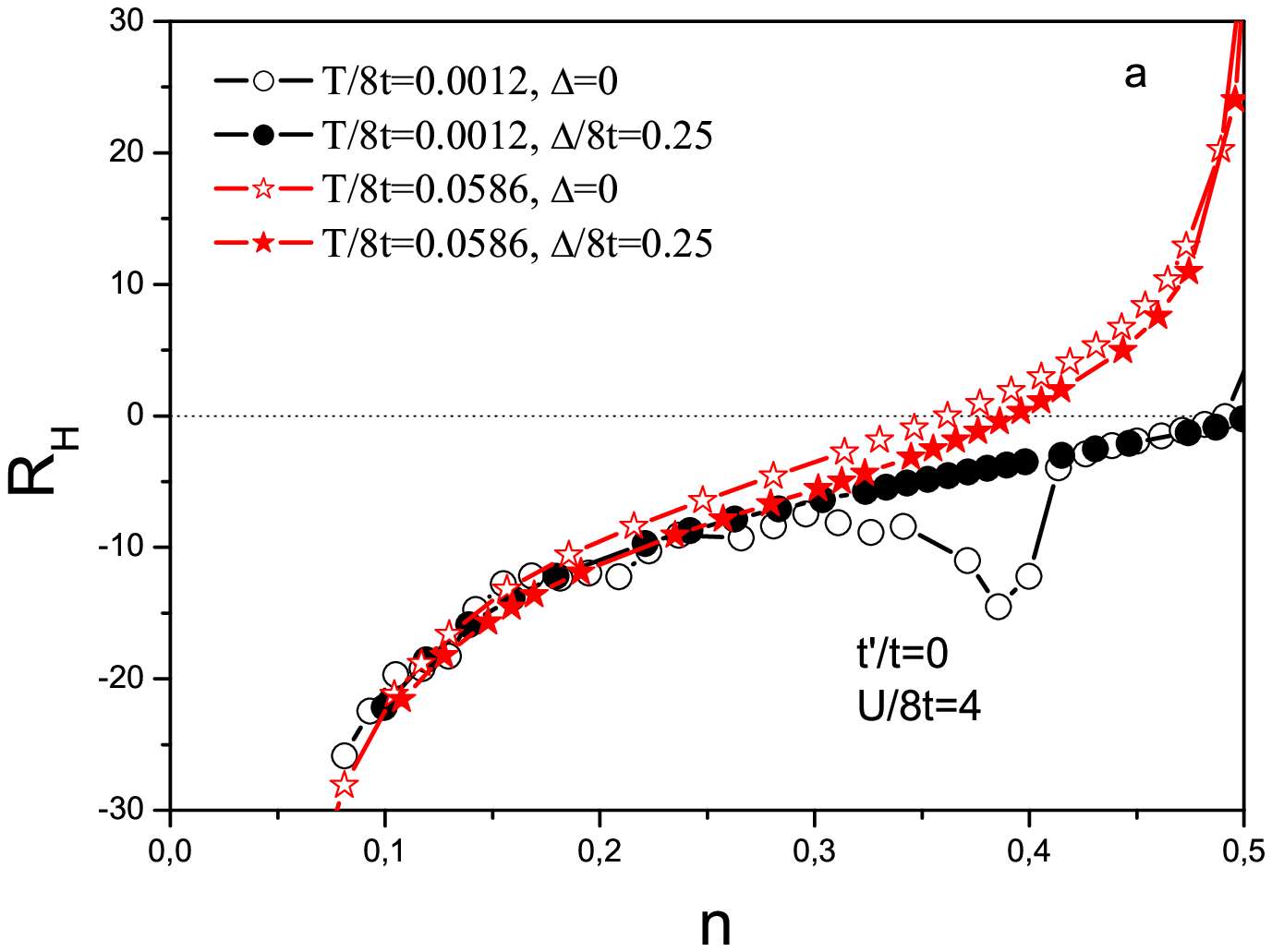}
\includegraphics[clip=true,width=0.5\textwidth]{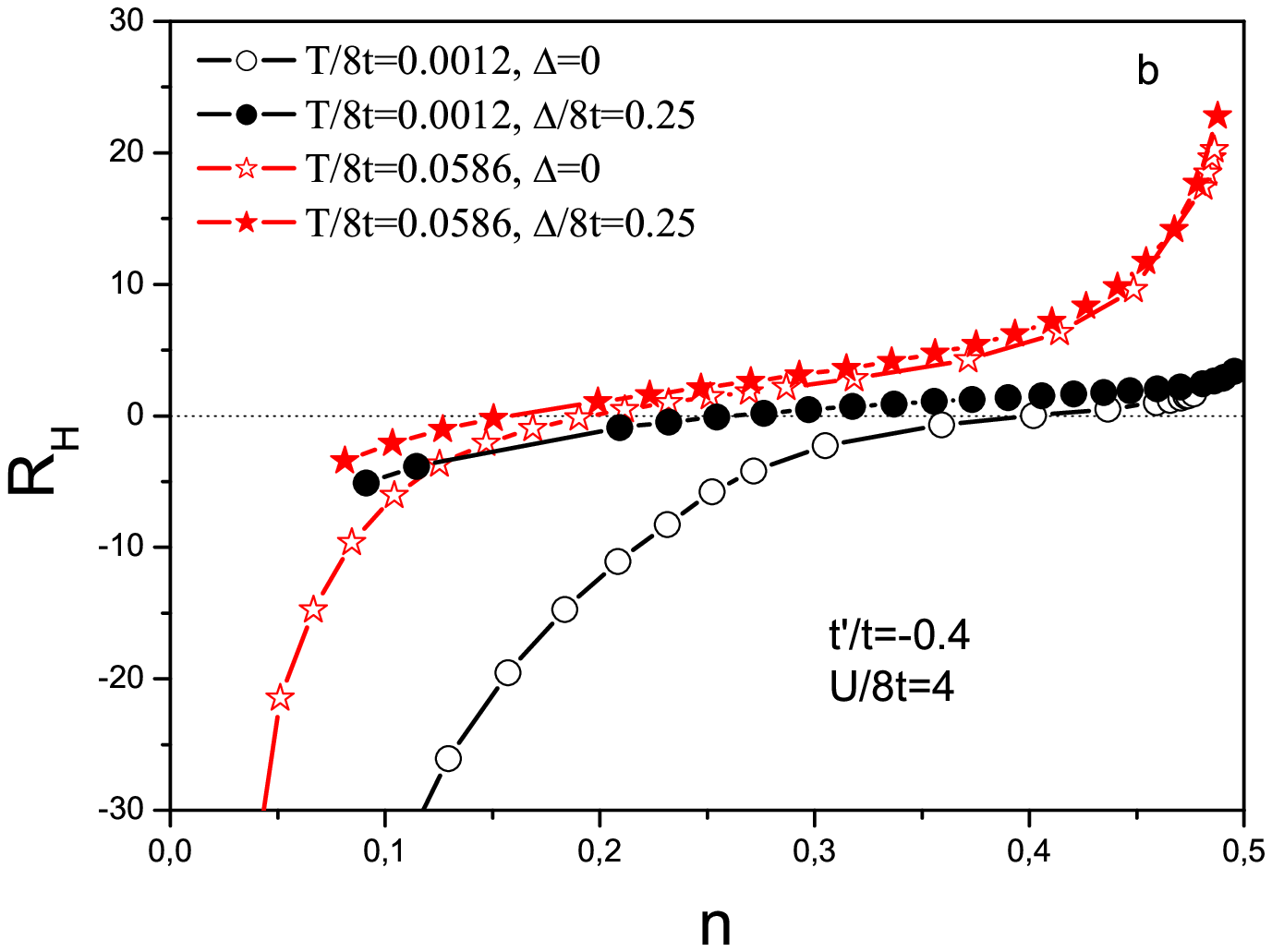}
\caption{Dependence of Hall coefficient on band filling in presence of impurity
scattering ($\Delta/8t=0.25$, filled symbols) and in its absence 
($\Delta=0$, empty symbols) for two models of two -- dimensional electronic
spectrum:
(a) full electron -- hole symmetry ($t'=0$); (b) $t'/t=-0.4$.}
\label{fig3}
\end{figure}

It should be noted that damping and disappearance of quasiparticle peak can
be not only due increasing temperature, but also due to disordering
\cite{GDMFT,dis_hubb_2008} (cf. Fig. \ref{fig3}) or due to pseudogap 
fluctuations, which are entirely neglected within local DMFT
\cite{GDMFT,DMFT+S}. Thus, in reality the region of applicability of simplest
estimates given above may be much wider.

In general case taking into account disorder scattering (more so pseudogap
fluctuations) in calculations of the Hall effect is rather complicated problem.
As a simple estimate we present below results of calculations using
Eqs. (\ref{Gxx}), (\ref{Gxy}), (\ref{R_H}), where we have used the values of 
the spectral density $A({\bf p}\varepsilon)$ for disordered Hubbard model
obtained within DMFT+$\Sigma$ approach \cite{GDMFT,DMFT+S}. Disorder parameter
$\Delta$ denotes the effective scattering rate of electrons by random field
(in self -- consistent Born approximation). It is clear that this approach 
based only on the account of disorder in spectral density is oversimplified,
but it seems reasonable for qualitative analysis.

In Fig. \ref{fig3} we compare the dependencies of Hall coefficient on band 
filling in the absence of disorder and for the case of impurity scattering with
$\Delta/8t=0.25$ for Mott insulator with $U/8t=4$. It is seen that
for different values of $t'$ in high temperature limit disorder only slightly
affects the Hall coefficient by rather insignificant shift of the value of
filling, where $R_H$ changes its sign. In low temperature regime impurity
scattering, damping the quasiparticle peak, lead to disappearance of the
anomalies of $R_H$, connected with its existence (cf. Fig. \ref{fig3}(a)) 
and weakening differences between low temperature and high temperature regimes.

\begin{figure}
\includegraphics[clip=true,width=0.5\textwidth]{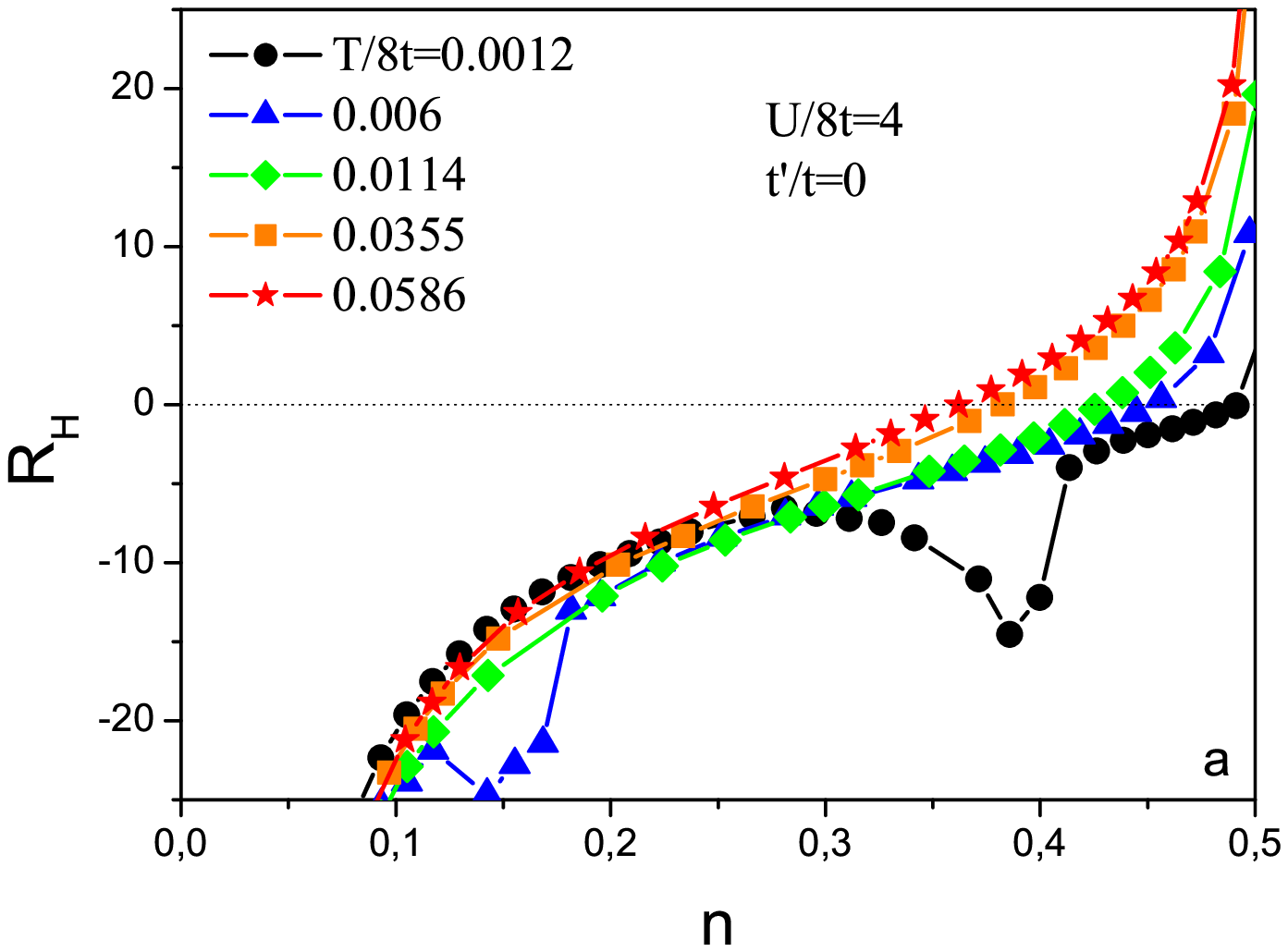}
\includegraphics[clip=true,width=0.5\textwidth]{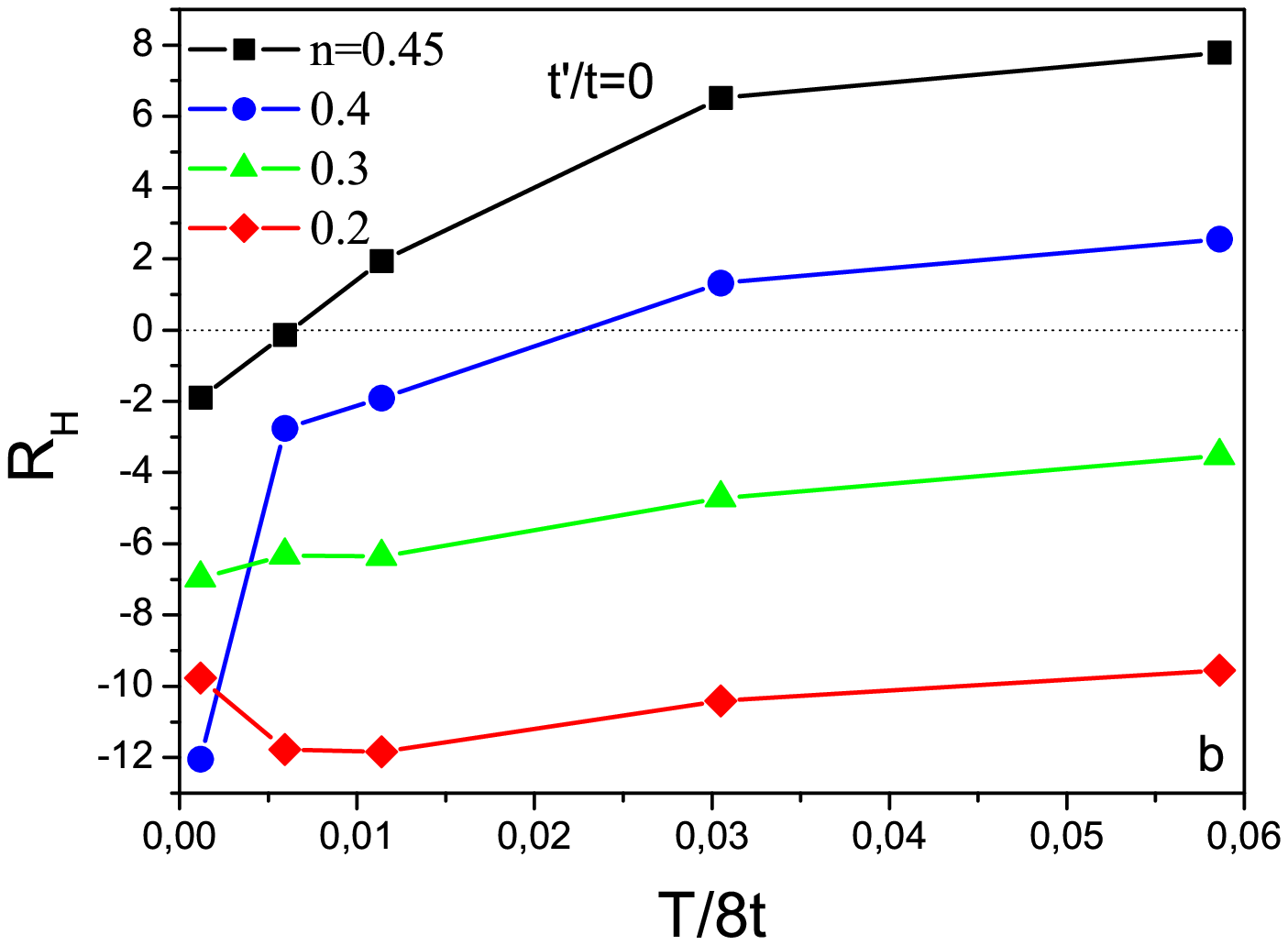}
\includegraphics[clip=true,width=0.5\textwidth]{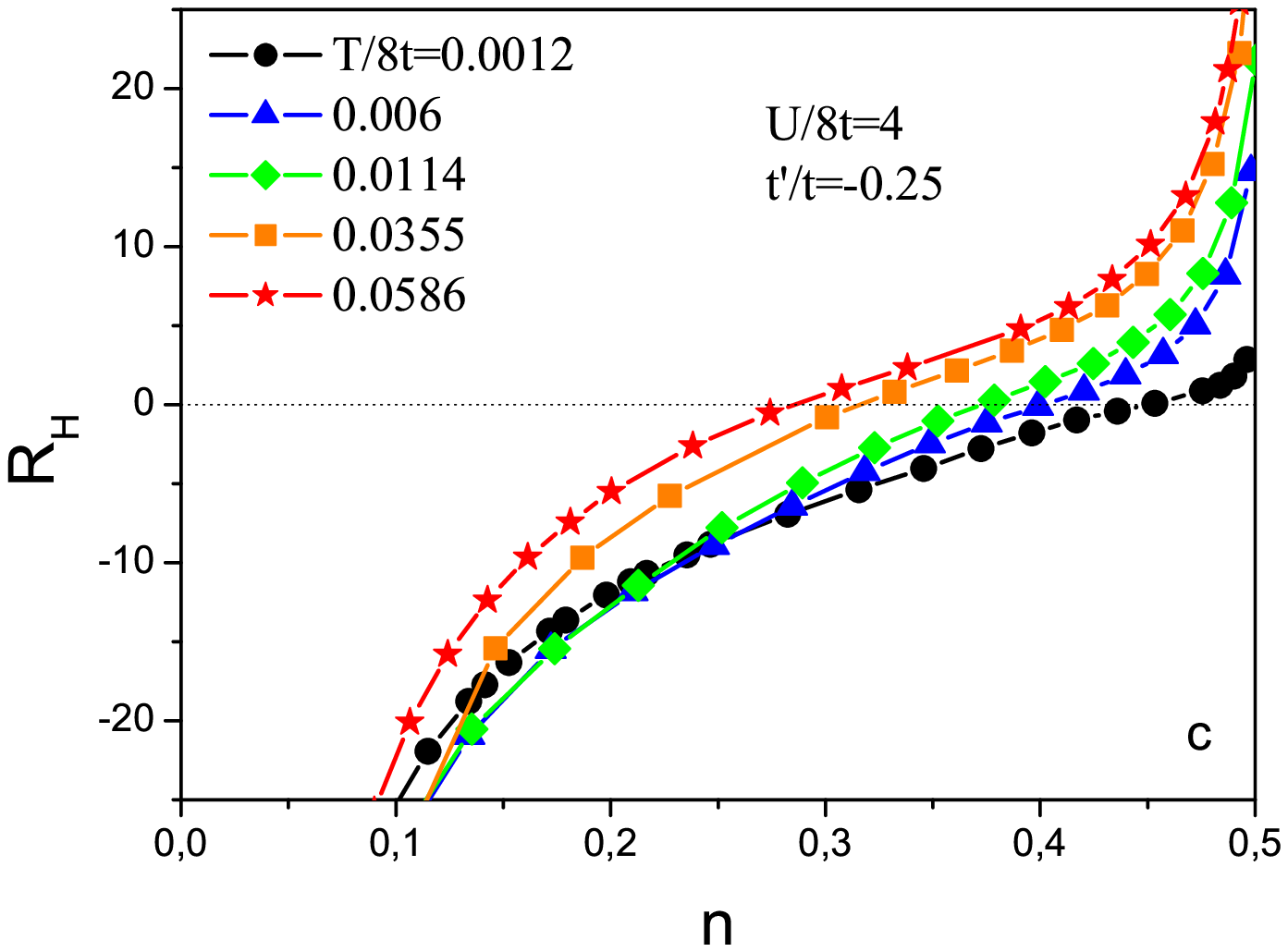}
\includegraphics[clip=true,width=0.5\textwidth]{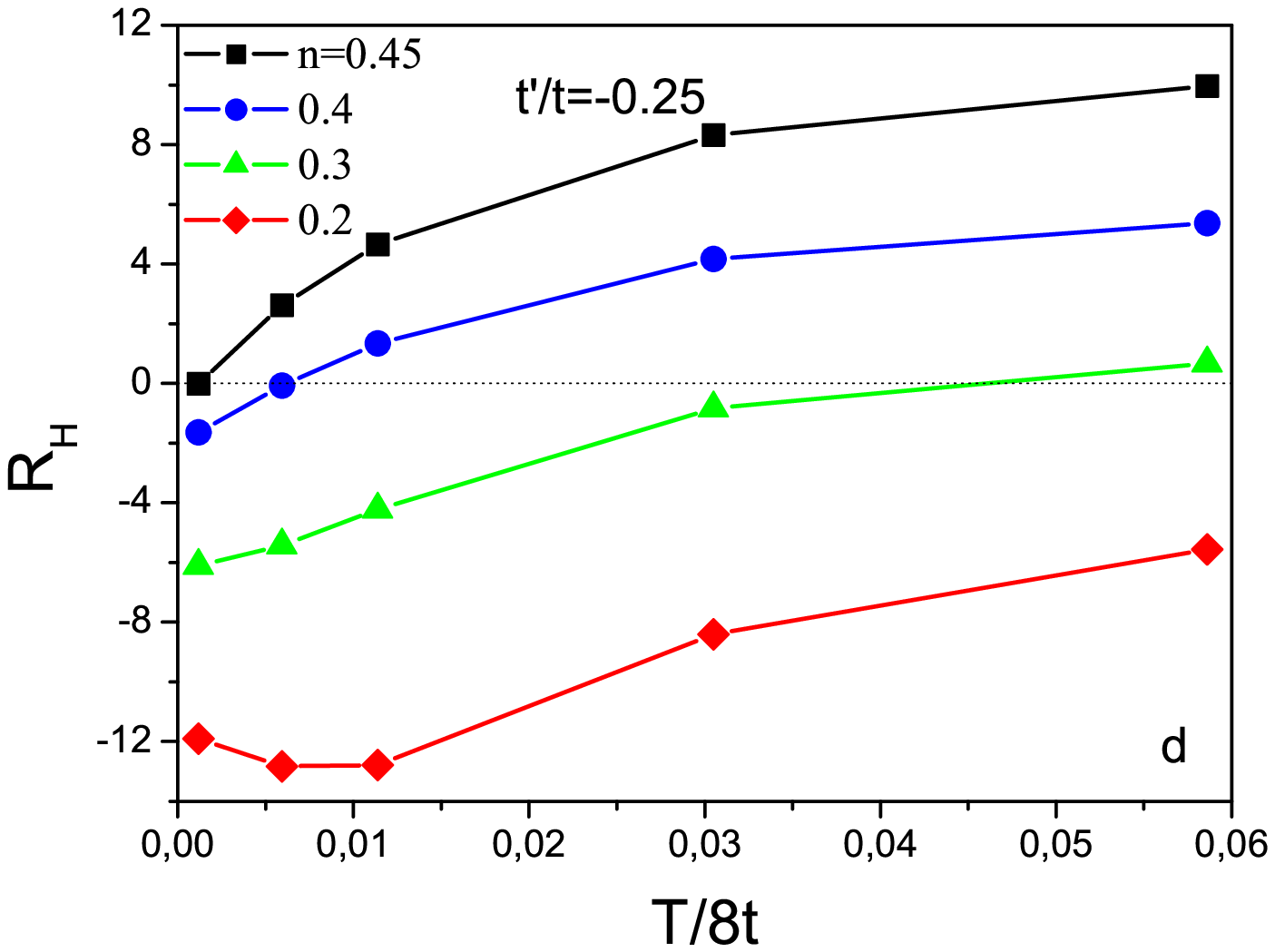}
\includegraphics[clip=true,width=0.5\textwidth]{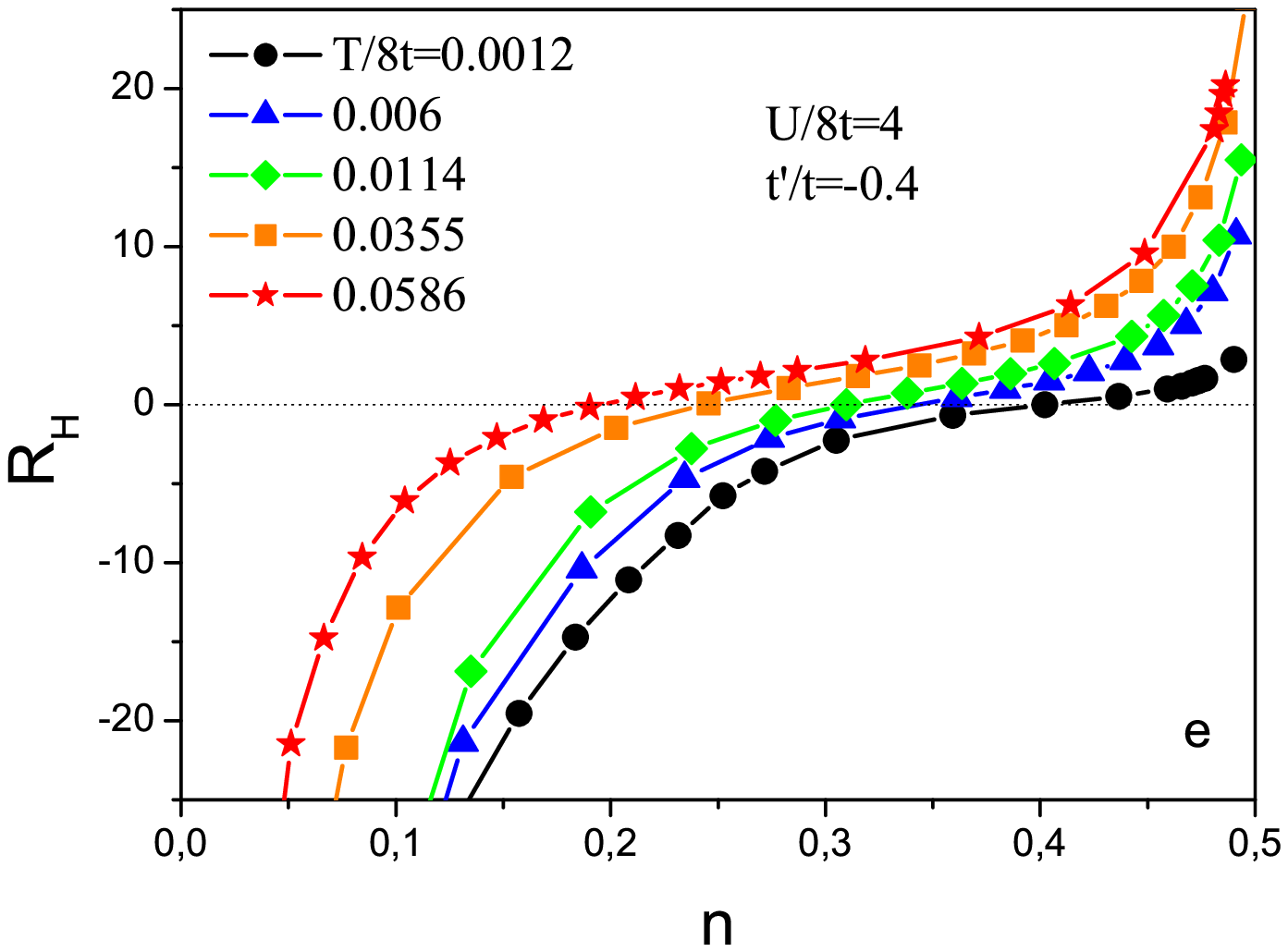}
\includegraphics[clip=true,width=0.5\textwidth]{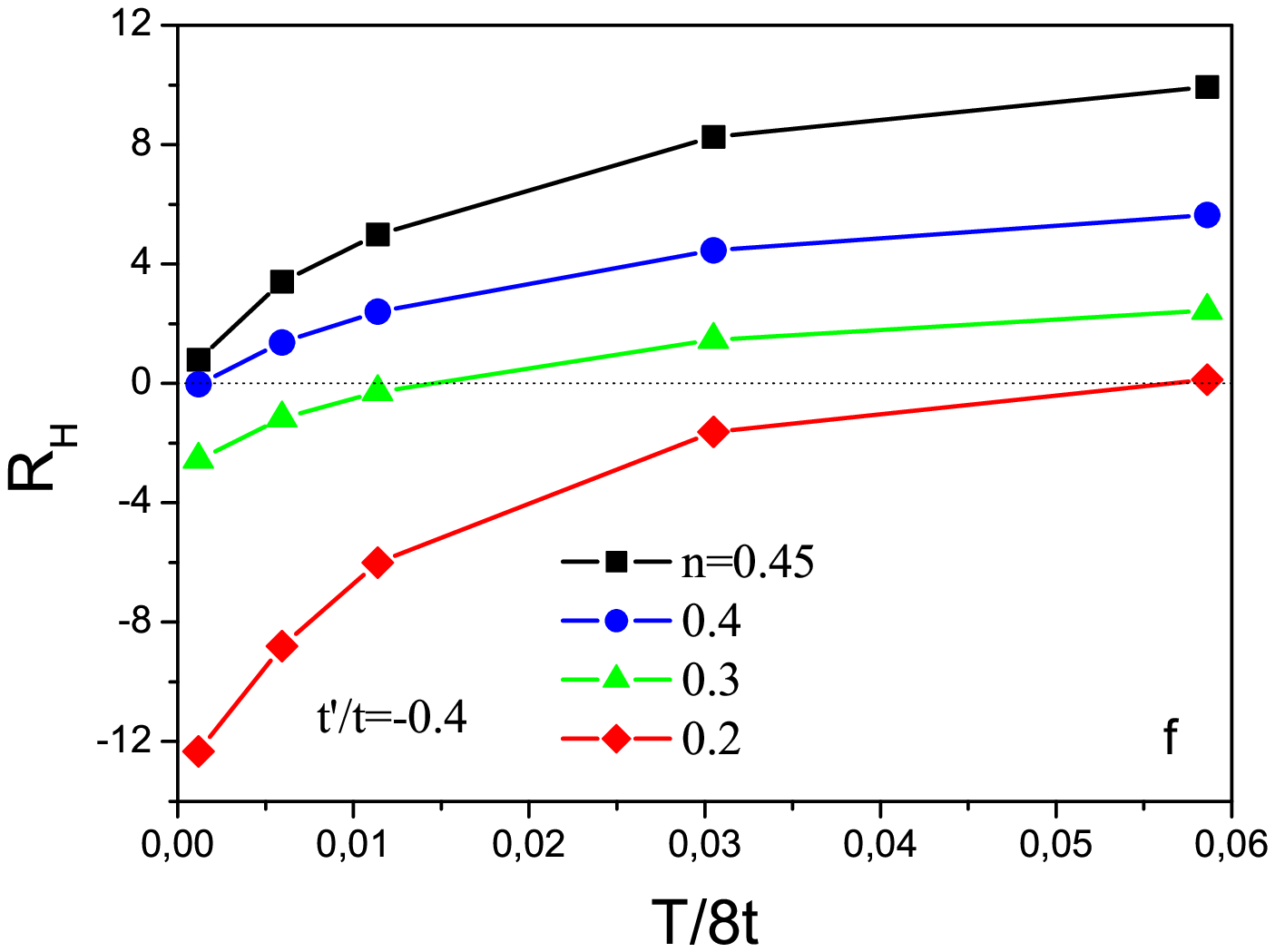}
\caption{Dependence of Hall coefficient on band filling for different values
of temperature -- left column (Fig. a,c,e) and temperature dependence of $R_H$ 
for different band fillings -- right column (Fig. b,d,f).}
\label{fig4}
\end{figure}

In Fig. \ref{fig4} we show dependencies of Hall coefficient on band filling and
temperature for the case of Mott insulator with $U/8t=4$ for different models
of electronic spectrum, both for the case of full electron -- hole symmetry
with $t'=0$ and for $t'/t=-0.25$ and $t'/t=-0.4$, characteristic for cuprate
systems LSCO and YBCO respectively. On the dependence of $R_H$ on band filling
with the growth of temperature we observe smooth evolution from low to high
temperature regime with smooth weakening of the anomalies of Hall coefficient
related to quasiparticle peak, which are most clearly seen in Fig. \ref{fig2}(a) 
and Fig. \ref{fig4}(a). For all cases of electronic spectrum under consideration
($t'/t=0;-0.25;-0.4$) increasing temperature leads to a shift of the value of
filling corresponding to $R_H=0$ into the region of larger hole dopings.

Also in the r.h.s. part of Fig. \ref{fig4}(b,d,f) we show the temperature 
dependencies of Hall coefficient for different band fillings.
In all case we observe the significant dependence of $R_H$ on temperature and
for small hole dopings ($n=0.45-0.3$) $R_H$ grows with increasing temperature
and we obtain the sign change of $R_H$ at larger hole dopings ($n=0.3-0.2$).
For small enough values of $t'$ ($t'/t=0;-0.25$) we can observe non monotonous
dependence of Hall coefficient on temperature, when $R_H$ decreases with 
increasing temperature, while it grows at high $T$.

\begin{figure}
\includegraphics[clip=true,width=0.5\textwidth]{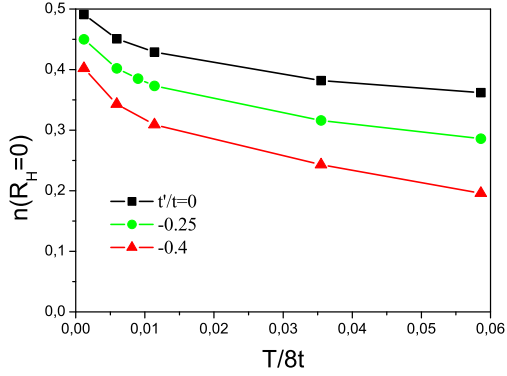}
\caption{Temperature dependence of band filling corresponding to a sign change
of Hall coefficient in doped Mott insulator for three different values of
$t'/t$.}
\label{fig5}
\end{figure}

The sign change of Hall coefficient is usually connected to a change of the type
of charge carriers. Hall coefficient approaching zero corresponds to
divergence of Hall number $n_{H}\sim 1/R_{H}$. At Fig. \ref{fig5} 
we show temperature dependence of the band filling corresponding to the sign
change of the Hall coefficient for all three values of $t'/t$ considered here. 
We see that in all models the band filling at which $R_H$ changes its sign
decreases with temperature. In case of the full electron -- hole symmetry
$t'=0$ we see, that in high temperature regime hole doping
$\delta=1-2n$ corresponding to the sign change of $R_H$ really tends to $1/3$. 
However, with increasing $|t'/t|$ we observe the significant decrease of the
value of hole doping where $R_H$ changes its sign.

\section{Hall coefficient in the model with semi -- elliptic density of states}

Let us briefly discuss results obtained in the model of electronic band with
semi -- elliptic density of states, which has the full electron -- hole
symmetry. The main results are qualitatively similar to the case two --
dimensional tight -- binding electronic spectrum with $t'=0$ also having the
complete electron -- hole symmetry. Similarly to two -- dimensional case the
Hall coefficient in three -- dimensional strongly correlated system is
significantly dependent on temperature and it is necessary to consider
separately the low and high temperature regimes for $R_H$, as in the low
temperature regime the Hall coefficient is mainly determined on filling the
quasiparticle band (quasiparticle peak).

\begin{figure}
\includegraphics[clip=true,width=0.6\textwidth]{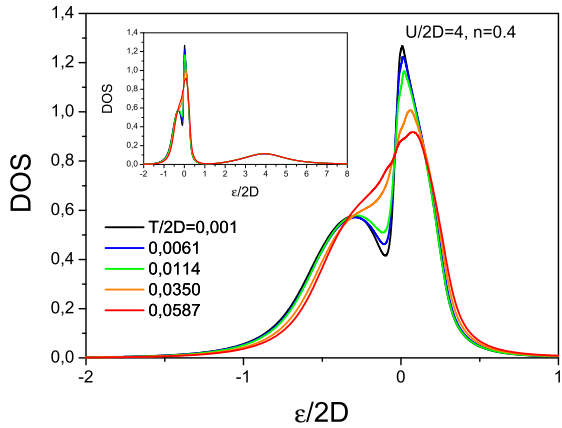}
\caption{Density of states (DOS) in doped Mott insulator with semi -- elliptic
band (three -- dimensional case) for different temperatures.}
\label{fig7}
\end{figure}

Increasing temperature leads to damping of quasiparticle peak
(cf. Fig. \ref{fig7}) and in high temperature regime Hall coefficient is
mainly determined by filling of the lower (for the case of hole doping 
considered here) Hubbard band.

\begin{figure}
\includegraphics[clip=true,width=0.5\textwidth]{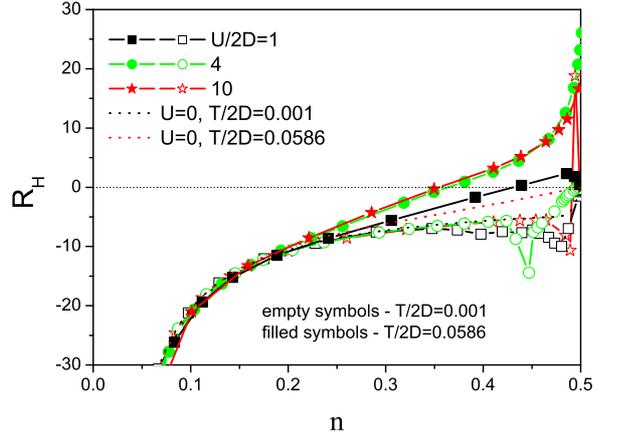}
\includegraphics[clip=true,width=0.5\textwidth]{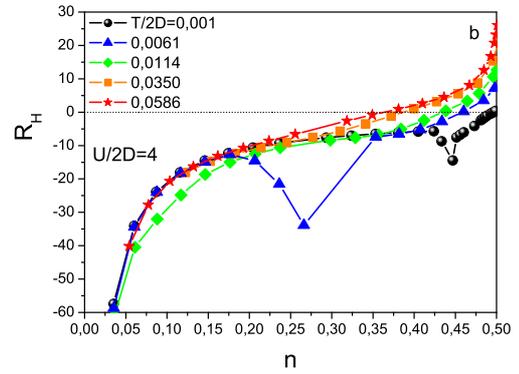}
\caption{Dependence of Hall coefficient on band filling for semi -- elliptic
density of states: (a) -- for different values of $U$
in low temperature (empty symbols) and high temperature (filled symbols)
regimes; (b) -- for different temperatures at fixed $U/2D=4$.}
\label{fig8}
\end{figure}

In Fig. \ref{fig8} (a) we show the Hall coefficient dependence on electronic 
band filling if low temperature (unfilled symbols) and in high temperature
(filled symbols) regimes, both for the case of strongly correlated metal
($U/2D=1$) and for doped Mott insulator ($U/2D=4;10$). We can see that in the
low temperature regime, as in two dimensional model with $t'=0$ ($R_H$ is
negative practically at all band fillings) at small hole dopings there is a
significant non monotonous dependence of $R_H$ on doping.

In high temperature regime the Hall coefficient at small hole dopings is
positive (hole -- like), decreasing with increasing hole doping, while at
larger dopings $R_H$ becomes negative, changing its sign (in Mott insulator)
at hole doping $\delta=1-2n\approx 1/3$, which again confirms qualitative
estimates given above. A smooth evolution of Hall coefficient dependence of
filling as temperature increases from low temperature to high temperature
regime in Mott insulator ($U/2D=4$) is shown in Fig. \ref{fig8}(b).

\onecolumn

\twocolumn

\begin{figure}
\includegraphics[clip=true,width=0.5\textwidth]{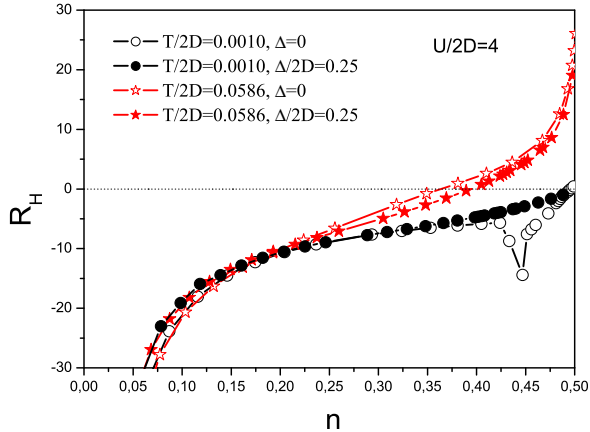}
\caption{Dependence of Hall coefficient on band filling in low temperature
regime (black curves) and high temperature regime (red curves)
in the absence of disorder $\Delta=0$ (empty symbols) and for $\Delta/2D=0.25$ 
(filled symbols).}
\label{fig9}
\end{figure}

In Fig. \ref{fig9} we demonstrate disorder influence on Hall coefficient in
Mott insulator. In high temperature limit impurity scattering practically does
not influence $R_H$ at all, while in low temperature limit damping the
quasiparticle peak by disorder removes the anomalous non monotonous behavior
of $R_H$ dependence on $n$.

\section{Comparison with experiments}

As we mentioned before in recent years the unique experimental studies were 
performed measuring Hall effect at low temperatures in the normal state of
high -- temperature  superconductors (cuprates), which was achieved in very
strong external magnetic fields\cite{Boeb,Tal1,Tal2}.
These experiments revealed the dependence of Hall number 
$n_H=\frac{a^2}{|eR_H|}$ on doping with a smooth transition from linear
dependence on hole concentration $\sim\delta$ at small dopings to the values
$\sim(1+\delta)$  for high enough concentrations of the order of critical 
hole concentration of vanishing (closing) pseudogap.
These data are usually interpreted within the picture of Fermi surface 
reconstruction in the vicinity of the expected quantum critical point in the
framework of rather specific model of cuprates with inhomogeneous localization
of carriers \cite{PrTal,Bar}. It should be noted, that in none of papers known
to us were in fact presented experimental points reliably demonstrating the
dependence $\sim(1+\delta)$, and clearly established experimental fact is only
the observed growth of the Hall number.

Below we propose an alternative interpretation of the growth of Hall number
in these experiments as reflecting the approach of the system to critical
concentration of carriers at which Hall effect just changes its sign
(Hall coefficient $R_H$ becomes zero) \cite{KKKS}.

\begin{figure}
\includegraphics[clip=true,width=0.5\textwidth]{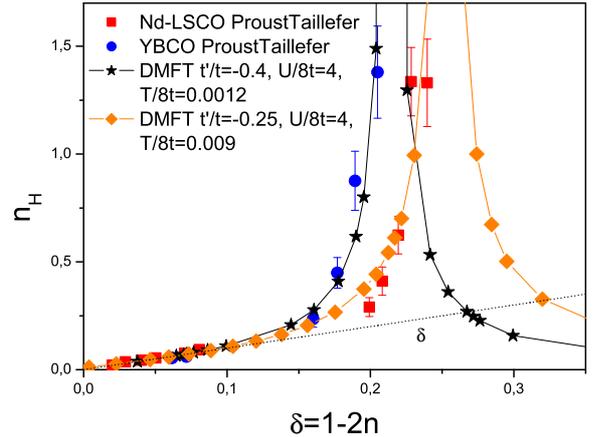}
\caption{Dependence of Hall number $n_H$ on doping -- comparison with
experiments \cite{Tal1,Tal2} on YBCO and Nd-LSCO, $\delta=1-2n$ -- hole
concentration. Stars and diamonds -- results of our calculations, blue circles
and red squares -- experiment.}
\label{fig6}
\end{figure} 

In Fig. \ref{fig6} we show the comparison of the results of our calculations
for Hall number (Hall concentration) $n_H=\frac{a^2}{|eR_H|}$ for typical
parameters of the model with experimental data for YBCO and Nd-LSCO
from Refs. \cite{Tal1,Tal2}. We can see that even for this, rather arbitrary,
choice of parameters we can obtain almost quantitative agreement with experiment,
with no assumptions about the connection of Hall effect with reconstruction of
Fermi surface by pseudogap and closeness to corresponding quantum critical
point, which were used in Refs. \cite{Tal1,Tal2,PrTal,Bar}.
Thus it seems reasonable to interpret Hall effect in cuprates within picture of
lower Hubbard model doping in Mott insulator, as an alternative to the scenario
based upon closeness to a quantum critical point. 

In this respect it seems to be quite important to try to perform more detailed
studies of the Hall effect in the vicinity of a critical concentration
corresponding to sign change of the Hall effect (divergence of the Hall number).
It requires the studies of systems (cuprates) where such sign change can be
achieved under doping.

\section{Conclusion}

We have studied Hall effect in metallic state appearing while doping Mott
insulator. Main attention was to the case of hole doping, characteristic for
a major part of cuprates. We considered a number of two -- dimensional tight --
binding models of electronic spectrum appropriate for description of
electronic structure of cuprates, as well as three -- dimensional model with
semi -- elliptic bare density of states. In all models the Hall coefficient
$R_H$ in doped Mott insulator is significantly dependent on temperature.
In low temperature limit  $R_H$ is mainly determined by the filling of
quasiparticle peak, which may lead to non monotonous dependence of Hall
coefficient on doping. In high temperature limit, when quasiparticle peak is
essentially damped,  $R_H$ is mainly determined by filling of the lower
(for hole doping) Hubbard band. In this limit the sign change of the Hall 
coefficient and corresponding divergence of the Hall number takes place,
in the simplest (symmetric) case close to the band filling 
$n$=1/3 per single spin projection or 2/3 for total density of electrons,
which corresponds to hole doping $\delta=1-2n$=1/3, though in general case
this filling may strongly depend on the choice of parameters of the model.
This concentration follows from simple qualitative estimates and not related
with more complicated factors like changing the topology of Fermi surface or
the presence of quantum critical points.

Rather satisfactory agreement of obtained concentration dependencies of Hall
number with experiments on YBCO and Nd-LS CO \cite{Tal1,Tal2} shows, that
our model may serve as a reasonable alternative to a picture of Hall effect
in the vicinity of quantum critical point related to closing the pseudogap
\cite{PrTal,Bar}.

The work of EZK, NAK and MVS was supported in part by RFBR grant 
No. 20-02-00011.  DIK work was partly supported by DFG project. 
No. 277146847 -- CRC 1238.


\vskip 1cm


\begin{center}

{\bf APPENDIX}  

\end{center}

\vskip 0.5cm

{\bf ``Bare'' electronic dispersion and its derivatives for band with
semi -- elliptic density of stetes.}

\vskip 0.5cm

Let us assume that electronic spectrum corresponding to density of ststes
(\ref{DOSd3}) is isotropic
$\varepsilon ({\bf p})=\varepsilon (|{\bf p}|)\equiv\varepsilon (p)$.
To calculate derivatives in (\ref{Gxx}) and (\ref{Gxy}) it is necessary to perform
``angle'' averaging of these derivatives by momentum components
\begin{equation}
\left<\left( \frac{\partial\varepsilon({\bf p})}{\partial p_x}
\right)^2\right>_\Omega=
\varepsilon '^2(p)\left<\frac{p_x^2}{p^2}\right>_\Omega 
=\frac{1}{d}\varepsilon '^2(p)= \frac{1}{3}\varepsilon '^2(p)
\label{e'2}
\end{equation}
where $<...>_\Omega=\int \frac{d\Omega}{4\pi}...$ is solid angle averaging
in three -- dimensional system ($d=3$) and $\varepsilon '(p)=
\frac{d\varepsilon (p)}{dp}$ is derivative over the absolute value of momentum.
\begin{eqnarray}
\left( \frac{\partial\varepsilon({\bf p})}{\partial p_x} \right) ^2 
\frac{\partial ^2\varepsilon({\bf p})}{\partial p_y^2}=\nonumber\\
=\varepsilon '^2(p) \left[ 
\varepsilon ''(p)\frac{p^2_x p^2_y}{p^4}+\frac{\varepsilon '(p)}{p}
\frac{p^2_x p^2-p^2_x p^2_y}{p^4}
\right]
\end{eqnarray}
where $\varepsilon ''(p)=\frac{d^2\varepsilon (p)}{dp^2}$. 
Thus we have a problem of finding the angle average 
$\left<\frac{p^2_x p^2_y}{p^4}\right>_\Omega$.
Let us introduce notations: $\left<\frac{p^4_x}{p^4}\right>_\Omega\equiv a$ 
and
$\left<\frac{p^2_x p^2_y}{p^4}\right>_\Omega \equiv b$. 
First of al we have:
\begin{eqnarray}
\left<\frac{(p^2_x+p^2_y+p^2_z)^2}{p^4}\right>_\Omega=\nonumber\\
=\left<\frac{(p^4_x+p^4_y+p^4_z)^2+2p^2_x p^2_y+2p^2_x p^2_z+2p^2_y p^2_z}{p^4}
\right>_\Omega=\nonumber\\
=d\left<\frac{p^4_x}{p^4}\right>_\Omega+d(d-1)\left<\frac{p^2_x p^2_y}{p^4}
\right>_\Omega=3a+6b=1.\
\label{ab1}
\end{eqnarray}
Similarly:
\begin{equation}
\left<\frac{(p^2_x+p^2_y)^2}{p^4}\right>_
\Omega=\left<\frac{p^4_x+p^4_y+2p^2_x p^2_y}{p^4}\right>_\Omega=
2a+2b=\frac{8}{15}
\label{ab2}
\end{equation}
As 
\begin{eqnarray}
\left<\frac{(p^2_x+p^2_y)^2}{p^4}\right>_\Omega =<sin^4\theta >_\Omega =
\nonumber\\
=\frac{1}{2}\int_{0}^{\pi}sin\theta sin^4\theta d\theta =
\frac{1}{2}\int_{-1}^{1}(1-\tau ^2)^2d\tau=\frac{8}{15}
\end{eqnarray}
where $\theta$ is an angle between vector ${\bf p}$ and $z$ -- axis, Then
from Eqs. (\ref{ab1}), (\ref{ab2}) we immediately obtain
$a=\left<\frac{p^4_x}{p^4}\right>_\Omega =1/5$ and
$b=\left<\frac{p^2_x p^2_y}{p^4}\right>_\Omega =1/15$, so that we have:
\begin{equation}
\left<\left( \frac{\partial\varepsilon({\bf p})}{\partial p_x} \right) ^2 
\frac{\partial ^2\varepsilon({\bf p})}{\partial p_y^2}\right>_\Omega =
\frac{\varepsilon '^2(p)}{15}[\varepsilon ''(p)+4\varepsilon '(p)/p]
\label{e'2e''}
\end{equation}

To find derivatives $\varepsilon '(p)$, $\varepsilon ''(p)$ for the
spectrum determined by semi-elliptic density of states (\ref{DOSd3}) we can 
use the approach developed in Ref.  \cite{dis_hubb_2008}.
Equating the number of states in a phase volume element  $d^3p$
and the number of states in an energy interval 
$[\varepsilon ,\varepsilon +d\varepsilon ]$], we obtain
differential equation determining $\varepsilon (p)$:
\begin{equation}
\frac{4\pi p^2dp}{(2\pi)^3}=N_{0}(\varepsilon )d\varepsilon
\label{DifEq1}
\end{equation}
Assuming the quadratic dispersion of $\varepsilon (p)$ close to lower
band edge we obtain the initial condition for (\ref{DifEq1}): $p\to 0$ for 
$\varepsilon\to -D$.
As a result:
\begin{equation}
p={\left[6\pi\left(\pi-\varphi +
\frac{1}{2}sin(2\varphi )\right)\right]}^{\frac{1}{3}}
\label{spektr1}
\end{equation}
where $\varphi =arccos(\frac{\varepsilon}{D})$ and momentum is given 
in units of inverse lattice parameter. This expression implicitly defines
the dispersion law $\varepsilon (p)$ on electronic branch of the spectrum
$\varepsilon\in [-D,0]$.

We can determine characteristic momentum $p_0$ corresponding to 
$\varepsilon =0$:
\begin{equation}
p_0=p(\varepsilon =0)={\left(3\pi^2\right)}^{\frac{1}{3}}
\label{p0}
\end{equation}
We are interested in calculating two derivatives of this spectrum over the
momentum. From (\ref{DifEq1}) we get:
\begin{equation}
\varepsilon '(p)=\frac{d\varepsilon}{dp}=\frac{p^2}{2\pi^2}\frac{1}
{N_0(\varepsilon )},
\label{veloc_e1}
\end{equation}
where $p$ is defined by (\ref{spektr1}).
\begin{eqnarray}
\varepsilon ''(p)=\frac{d}{dp}\frac{d\varepsilon}{dp}=
\frac{1}{2\pi^2}\frac{2pN_0(\varepsilon )-p^2\frac{dN_0(\varepsilon )}
{d\varepsilon}
\frac{d\varepsilon}{dp}}{N^2_0(\varepsilon )}=\nonumber\\
=\frac{1}{N_0(\varepsilon )}
\left[ \frac{p}{\pi^2}-\varepsilon '^2(p)\frac{dN_0(\varepsilon )}{d\varepsilon}
\right]
\label{e''}
\end{eqnarray}
where $\frac{dN_0(\varepsilon )}{d\varepsilon}=-\frac{2}{\pi D^2}\frac{\varepsilon}
{\sqrt{D^2-\varepsilon^2}}$, $\varepsilon '(p)$ is determined from 
(\ref{veloc_e1}), while $p$ is defined by (\ref{spektr1}).

On the hole branch of the spectrum ($\varepsilon\in [0,D]$), to obtain quadratic
dispersion law close to the upper edge of the band ($\varepsilon\to D$) we 
introduce a hole momentum $\tilde p=2p_0-p$  and equate the number of states 
in a phase volume element $d^3\tilde p$ and in energy interval
$[\varepsilon ,\varepsilon +d\varepsilon]$:
\begin{equation}
\frac{4\pi\tilde p^2d\tilde p}{(2\pi)^3}=-N_{0}(\varepsilon )d\varepsilon
\label{DifEq2}
\end{equation}
Demanding $\tilde p\to 0$ at the upper band edge $\varepsilon\to 0$, we obtain:
\begin{equation}
\tilde p={\left[6\pi\left(\varphi -\frac{1}{2}sin(2\varphi )\right)
\right]}^{\frac{1}{3}}
\label{spektr2}
\end{equation}
For the velocity on the hole branch of the spectrum we get:
\begin{equation}
\varepsilon '(p)=-\frac{d\varepsilon}{d\tilde p}=
\frac{\tilde p^2}{2\pi^2}\frac{1}{N_0(\varepsilon )}
\label{veloc_e2}
\end{equation}
Eqs. (\ref{veloc_e1}), (\ref{veloc_e2}) determine the dependence of velocity
$\varepsilon '(p)$ on energy. One is easily convinced that velocity is even
in energy and goes to zero at band edges. The second derivative over momentum
in this approach is explicitly defined on electronic branch of the spectrum
($\varepsilon\in [-D,0]$), but on the hole branch it is more difficult to do.
However, we can require full electron -- hole symmetry of the model, which
reduces to demanding the square of velocity, entering Eq. (\ref{Gxx}), 
being even in $\varepsilon (p)$, while Eq. (\ref{e'2e''}) entering
Eq. (\ref{Gxy}) for Hall conductivity being odd (sign change under change of
the type of charge carriers).With the account of such symmetry the results
obtained in this Appendix allow to replace summation over momenta in Eqs.
(\ref{Gxx}), (\ref{Gxy}) by integration over energy.


\end{document}